\documentclass[runningheads]{llncs}

\usepackage[mode=buildnew]{standalone}
\usepackage{acronym}
\usepackage{enumitem}
\usepackage{xtab}
\usepackage{tabularx}
\usepackage{url}
\usepackage[hidelinks]{hyperref}
\usepackage{booktabs}
\usepackage{wrapfig}

\usepackage[cachedir=resources/minted-cache,frozencache]{minted}
\usepackage{subfig}
\usepackage[T1]{fontenc}
\usepackage[utf8]{inputenc}
\usepackage{multicol}
\usepackage{rotating}
\usepackage{pdflscape}
\usepackage{import}
\usepackage{pifont}
\DeclareUnicodeCharacter{2714}{\ding{52}}
\DeclareUnicodeCharacter{2718}{\ding{54}}
\DeclareUnicodeCharacter{25CF}{\ding{108}}
\DeclareUnicodeCharacter{274D}{\ding{109}}
\DeclareUnicodeCharacter{2798}{\ding{216}}
\DeclareUnicodeCharacter{279A}{\ding{218}}

\acrodef{ACS}{Auto Configuration Server}\acrodefplural{ACS}[ACSes]{Auto Configuration Servers}
\acrodef{DSL}{Digital Subscriber Line}\acused{DSL}
\acrodef{API}{Application Programming Interface}\acused{API}
\acrodef{CA}{Certificate Authority}\acrodefplural{CA}{Certificate Authorities}
\acrodef{CPE}{Customer Premises Equipment}
\acrodef{CWMP}{\acs{CPE} \acs{WAN} Management Protocol}
\acrodef{DHCP}{Dynamic Host Configuration Protocol}\acused{DHCP}
\acrodef{HTTP}{Hypertext Transfer Protocol}\acused{HTTP}
\acrodef{HTTPS}{\acl{HTTP} over \acs{TLS}}\acused{HTTPS}
\acrodef{IP}{Internet Protocol}\acused{IP}
\acrodef{LAN}{Local Area Network}\acused{LAN}
\acrodef{MAC}{Media Access Control}\acused{MAC}
\acrodef{RPC}{Remote Procedure Call}
\acrodef{SOAP}{Simple Object Access Protocol}\acused{SOAP}
\acrodef{SSL}{Secure Sockets Layer}\acused{SSL}
\acrodef{STUN}{Session Traversal Utilities for Network Address Translation}\acused{STUN}
\acrodef{TCP}{Transmission Control Protocol}\acused{TCP}
\acrodef{TLS}{Transport Layer Security}\acused{TLS}
\acrodef{TOR}{The Onion Router}\acused{TOR}
\acrodef{URL}{Uniform Resource Locator}\acused{URL}
\acrodef{USB}{Universal Serial Bus}\acused{USB}
\acrodef{VoIP}{Voice over \acs{IP}}%
\acrodef{WAN}{Wide Area Network}\acused{WAN}
\acrodef{XML}{Extensible Markup Language}\acused{XML}
\acrodef{XMPP}{Extensible Messaging and Presence Protocol}\acused{XMPP}

\expandafter\def\expandafter\UrlBreaks\expandafter{	\do\-\do\.\do\_}
\begin{document}
\title{Watching the Weak Link into Your Home:\\An Inspection and Monitoring Toolkit for TR-069}
\titlerunning{An Inspection and Monitoring Toolkit for TR-069}
\subtitle{Full Version}
\author{
	Maximilian Hils%
	\and 
	Rainer Böhme%
}
\authorrunning{M. Hils \and R. Böhme}
\institute{University of Innsbruck, Austria\\
	\email{\{maximilian.hils,rainer.boehme\}@uibk.ac.at}
}
\maketitle

\begin{abstract}
	TR-069 is a standard for the remote management of end-user devices by service providers. Despite being implemented in nearly a billion devices, almost no research has been published on the security and privacy aspects of TR-069.
	The first contribution of this paper is a study of the TR-069 ecosystem and techniques to inspect TR-069 communication. We find that the majority of analyzed providers do not use recommended security measures, such as TLS.
	Second, we present a TR-069 honeyclient to both analyze TR-069 behavior of providers and test configuration servers for security vulnerabilities. We find that popular open-source configuration servers use insecure methods to authenticate clients. TR-069 implementations based on these servers expose, for instance, their users' internet telephony credentials.
	Third, we develop components for a distributed system to continuously monitor activities in providers' TR-069 deployments. Our setup consists of inexpensive hardware sensors deployed on customer premises and centralized log collectors.
	We perform real-world measurements and find that  the purported security benefits of TR-069 are not realized as providers' firmware update processes are lacking.
	
	\keywords{TR-069, CWMP, ACS, Customer Premises Equipment, Remote Management, User Privacy, Monitoring, Honeyclient} 
\end{abstract}

\newcommand{\AcsTable}{Figure~\ref{fig:providerconfs}}
\newcommand{\CommandTable}{Appendix~\ref{sec:rpcs}}
\newcommand\blfootnote[1]{%
	\begingroup
	\renewcommand\thefootnote{}\footnote{#1}%
	\addtocounter{footnote}{-1}%
	\endgroup
}

\section{Introduction}
\vspace{-.35cm}
TR-069 is a technical specification that defines a protocol for the remote management of end-user devices by (internet) service providers. Published by the Broadband Forum, it is also endorsed by a variety of other initiatives (e.g.\ the DVB~Project and the WiMAX Forum). TR-069 is implemented in nearly a billion residential gateways, set-top boxes and other home broadband equipment~\cite{onebillion}.\blfootnote{Source Code Repository:\url{https://github.com/mhils/tr069}}

\sloppy
While TR-069 simplifies device setup for consumers, it provides service providers with unrestricted access to the managed device. This may compromise users' privacy and security expectations 
in a number of ways:
First, service providers can read and write any data on controlled devices. They can query routers for connected devices, obtain television viewing statistics from set-top boxes, or install new firmware. These operations are generally opaque to users and the protocol does not specify a way to seek users' consent.
Second, a configuration server can identify connected users and provide tailored firmware to individuals of interest. Providers may be coerced to help authorities to pivot into otherwise protected networks. 
Third, the configuration server used by a service provider to control TR-069 clients is a single point of failure. If compromised, an attacker gains control over all connected devices. 
Fourth, TR-069 devices accept HTTP requests on port 7547 in order to initiate sessions. This exposes them to direct attacks.

\vspace{-.2cm}
Arguably, TR-069's ubiquity and opaqueness make it an attractive platform for privacy-infringing data collection, attacks by criminals, or targeted surveillance.
For example, attacks on TR-069 devices have contributed to the quick growth of the Mirai botnet and caused widespread internet outages in Germany in 2016 \cite{understanding-mirai}.
Even though internet-wide scanning has shown that port 7547 is the third most exposed port on the internet \cite{scanning-liveness,censys15}, 
very little research has been published on the protocol's security. So far, researchers have primarily pointed out specific implementation vulnerabilities in clients and configuration servers \cite{defcon22shahar,31c3cooks}. This paper takes a broader look at the TR-069 ecosystem from a security and privacy perspective.

\vspace{-.2cm}
We will first give a security-centric overview of TR-069 in Section~\ref{sec:preliminaries}.
As the first contribution of this paper, we discuss different methods to inspect TR-069 connections and compare their practical feasibility in Section~\ref{sec:inspect}. This allows researchers and proficient users to view and analyze which data is communicated by TR-069-enabled devices in practice. We extract TR-069 configurations from firmware update packages and search them for TR-069 credentials. We find that the majority of providers do not follow recommended security practices.

\vspace{-.2cm}
In Section~\ref{sec:honeyclient}, we introduce an open-source TR-069 client that can be used to analyze configuration servers. %
By emulating a real TR-069 device, the client can determine %
which data is regularly accessed by the service provider during normal operation or which firmware is deployed to users.
This client can also be used to find security vulnerabilities in configuration servers. %
To demonstrate this application, we test two open-source configuration servers with our client. We find that both do not authenticate clients securely and can be tricked into exposing other users' credentials. As the issues we uncover are not specific to the individual implementations, our findings point to widespread issues regarding client authentication in the TR-069 landscape.

\vspace{-.2cm}
Looking at the implementation of TR-069 in practice, it is interesting to compare how different service providers make use of the protocol.
Section~\ref{sec:monitoring} presents a sensor software to monitor individual TR-069 devices for privacy infringements and other undesired actions. To make running the sensor practical, we show how a small and inexpensive router can be turned into a long-term TR-069 monitoring platform.
We perform measurements of real-world TR-069 traffic over a period of twelve months. We do not observe privacy violations by providers, but we find that the the purported security benefits of TR-069, such as the automated deployment of security updates, are not realized.

\vspace{-.2cm}
We discuss our findings in Section~\ref{sec:discussion}, review related work in Section~\ref{sec:related-work}, and conclude in Section~\ref{sec:conclusion}. Given the sensitive nature of TR-069 communications, we document our approach to research ethics and privacy in Appendix~\ref{sec:ethics}.\vspace*{-1.5cm}

\section{Preliminaries}
\label{sec:preliminaries}

TR-069 is a technical report that defines an application layer protocol for the remote management of end-user routers, modems, set-top boxes and other \ac{CPE} \cite{tr069}. It specifies generic methods that can be applied to manage different types of devices. Additional technical reports specify the data models for concrete device classes or services \cite{alltrs}.

TR-069 was initially published in 2004 by the Broadband Forum\footnote{
	The Broadband Forum is a non-profit industry consortium that defines standards for broadband networking. Members include internet and telecommunications service providers as well as device vendors.
} as a means to configure home routers and business internet gateways, but has since then evolved to cover \ac{VoIP} products, video set-top boxes, network attached storage, femto cells, and other networked products \cite{qacafeoverview}. 
TR-069 is endorsed by other initiatives, such as the Digital Video Broadcasting consortium%
and the WiMAX Forum. %
While the protocol defined in TR-069 is formally entitled \emph{\acf{CWMP}}, both terms are used interchangeably in practice and we will stick with the better known TR-069. %
Unless otherwise noted, this paper refers to TR-069 Amendment~6 (\ac{CWMP} Version~1.4) published in April 2018 \cite{tr069}.

In the following, we first describe the goals of TR-069 %
and then provide a technical introduction to the protocol.

\begin{figure}[t]
	\centering
	\includestandalone[page=1,width=.95\linewidth]{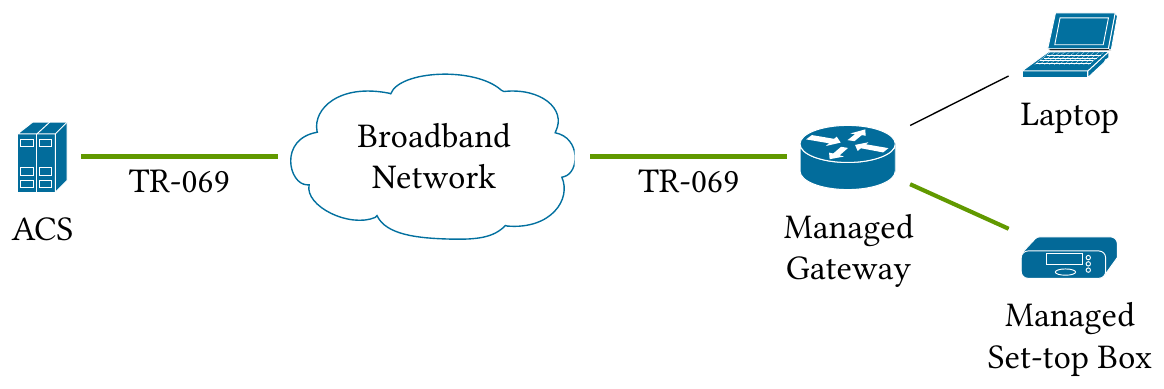}
	\vspace{-0.5cm}
	\caption{
		\label{fig:tr069network}
		Basic TR-069 network topology.
	}
	\vspace{-0.3cm}
\end{figure}

\subsection{TR-069 Goals}
\label{sec:preliminaries-tr069}
\vspace{-.2cm}

\sloppy{On a functional level, TR-069 covers a wide range of use cases for network providers. We briefly recall the key features to give readers a better understanding of TR-069 usage in practice.}

\begin{description}
	\item[``Zero-Touch'' Installation.] TR-069 is often used for auto-provisioning of new devices, which simplifies the setup procedure for new customers and thereby reduces support costs. Customers are provided with a modem or router that just needs to be connected to the network. It automatically contacts the provider's \ac{ACS} to receive its configuration (e.g.\ \ac{VoIP} credentials).
	\item[Firmware Maintenance.] Providers can deploy firmware updates to their customers without requiring user interaction.
	\item[Diagnostics for Customer Troubleshooting.] \sloppy{Customer service agents can remotely access diagnostic information and modify configuration settings of TR-069 devices to help users with troubleshooting their network.}
	\item[Configuration and Activation of New Services.] TR-069 can be used to remotely (de-)activate services or features on a customer's device. For example, some providers charge a monthly fee for activating the router's wireless module. Wireless network functionality is activated remotely via TR-069 when a customer subscribes to the service.

\end{description}

\noindent
Each of TR-069's use cases comes with its own set of security and privacy challenges. While TR-069 introduces elements that could potentially improve user security (e.g.\ automated firmware updates), we suppose that its adoption is primarily motivated by prospective reductions in service and support costs.

\subsection{TR-069 Protocol}

Here we provide a short introduction to the TR-069 protocol.
As the official specification counts 276 pages (excluding data models), we do not aim to provide a comprehensive overview but focus on the parts relevant for this paper. We assume the reader to be familiar with \ac{TCP}/\ac{IP}, \ac{TLS} and \ac{HTTP}.

\noindent
Fundamentally, TR-069 describes the interaction between an \acf{ACS} and a TR-069 client (hereinafter ``client''). Figure~\ref{fig:tr069network} shows an example topology: TR-069 communication usually happens only within a provider's network, with the client being located on customer premises and the \ac{ACS} being part of the service provider's infrastructure. However, some vendors also offer configuration servers as Software-as-a-Service in the cloud%
.

\begin{figure}[t]
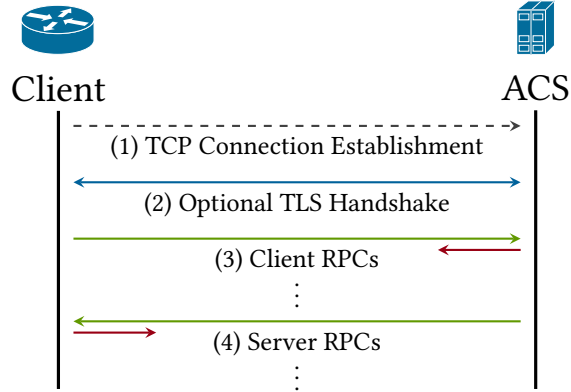

	\centering
	\includestandalone[page=2]{resources/illustrations}
	\caption{
		\label{fig:tr069handshake}
		High-level view on a TR-069 session.
	}
\end{figure}

\noindent
On a high level, connections follow the structure shown in Figure~\ref{fig:tr069handshake}. A TR-069 session is always initiated by the client, which establishes a \ac{TCP} connection with the \ac{ACS}. Second, the connection is optionally secured using \ac{TLS}. Third, the client sends a series of 
commands %
to the server. Finally, the \ac{ACS} sends a series of 
commands %
to the client before the connection is terminated. We discuss each step in more detail below.

\subsubsection{\acs{ACS} Discovery} 
\label{sec:acs-discovery}
Before making first contact with an \ac{ACS}, a client needs to learn the \ac{ACS} \ac{URL} it is supposed to contact. TR-069 defines three mechanisms for this: 
The use of a preconfigured \ac{ACS} \ac{URL}, configuration via a  \ac{LAN}-side protocol (e.g.\ a router web interface or TR-064 \cite{tr064}), or configuration via \ac{DHCP} options 43 and 125 during \ac{IP} address assignment. %

It is noteworthy that the \ac{ACS} \ac{URL} determines whether the connection will be secured using \ac{TLS}. While the TR-069 specification recommends the use of \ac{TLS}, it is not a mandatory part of the protocol. Therefore, control over the \ac{ACS} \ac{URL} can be used to downgrade the security of TR-069 sessions. In particular, the \ac{DHCP} discovery mechanism is prone to man-in-the-middle attacks unless otherwise secured \cite[p. 34]{tr069}. 

\subsubsection{Connection Establishment}
\label{sec:protocol-connection-establishment}
As the first step of a TR-069 session, the client opens a \ac{TCP} connection to the hostname specified in the configured \ac{ACS} \ac{URL}.
TR-069 sessions are always initiated by the client as a reaction to one or multiple events. Among others,\footnote{For a comprehensive listing of non-vendor-specific event types, \mbox{see \cite[p. 65]{tr069}}.} a connection may be initiated for the following reasons:
\newcommand{\ev}[2]{
	\textbf{#2}}
\begin{itemize}
	\item \ev{0}{Bootstrap}: Indicates that the session was established due to first-time client installation or a change to the \ac{ACS} \ac{URL}. The \ac{ACS} transmits initial configuration data to the client.
	\item \ev{2}{Periodic}: Indicates that the session was established on a periodic interval, e.g., once every 24 hours.
	\item \ev{4}{Value Change}: Indicates that the value of one or more monitored parameters has been modified.
	\item \ev{6}{Connection Request}: Indicates that the session is a reaction to a Connection Request from the \ac{ACS} (see below).
\end{itemize}

\vspace{-.2cm}
\noindent
After successful connection establishment, the client sends an \texttt{Inform} message which indicates the event type(s) that triggered the connection. This allows the \ac{ACS} to learn the client's intent and act accordingly.\vspace{-.4cm}

\subsubsection{Connection Requests}
For some of TR-069's use cases (e.g. live diagnostics, see Section \ref{sec:preliminaries-tr069}) it is necessary for the \ac{ACS} to request that the client contacts it immediately. TR-069 defines a mandatory Connection Request mechanism to provide this functionality.

The basic mechanism is a simple \ac{HTTP} GET request from the \ac{ACS} to the TR-069 client, which runs an \ac{HTTP} server for this purpose. The use of \ac{TLS} for the connection request is prohibited by the specification. To prevent denial-of-service attacks, \ac{HTTP} digest authentication is used to verify the validity of the request. Connection requests act as a trigger only: The client will connect back to the configured \ac{ACS} \ac{URL}, which cannot be modified by the connection request.

\newpage\noindent
A major limitation of the basic connection request mechanism is that the \ac{ACS} must be able to reach the client over \ac{HTTP}. While this was negligible when TR-069 was introduced for routers and gateways, it does not work for endpoints which reside behind firewalls or in (\ac{IP}v4) networks with network address translation. To accommodate for these devices, TR-069 Amendment~5 introduces an \ac{XMPP}-based mechanism, which supersedes the previously introduced \ac{STUN} tunneling in \mbox{TR-069} Annex G \cite{tr069}. In short, there are multiple methods the \ac{ACS} can use to instruct the client to contact it immediately. All mechanisms only serve as a trigger for a client-initiated TR-069 session. The client will always connect to the configured \ac{ACS} \ac{URL} for the actual TR-069 session.

While all approaches increase the attack surface, the basic connection request mechanism is particularly interesting to attackers for its public facing web server on the client. The connection request server listens on port 7547 by default, which is the third most publicly exposed port in the \ac{IP}v4 address space (after ports 80 and 443, before port 22) \cite{scanning-liveness,censys15}.

\vspace{-.2cm}
\subsubsection{Connection Security}
\label{sec:preliminaries-security}
If the \ac{ACS} \ac{URL} scheme is \ac{HTTPS}, the TR-069 session must be secured using \ac{TLS} before any messages are being sent. However, the specification imposes nonstandard restrictions on the use of \ac{TLS}, which may raise security concerns:

\begin{description}[leftmargin=.5cm]
	\item[Hostname Matching] The \ac{ACS} hostname must be validated against the certificate's Common Name, even though Common Name matching has been deprecated since \ac{HTTPS} was first introduced in 2000 in favor of Subject Alternative Names \cite{rfc2818}. TR-069 describes a simple -- yet nonstandard -- exact string matching method, which is not available in major \ac{TLS} libraries and must be implemented manually to fully comply with the specification.
	
	\item[Root Certificate Set] The client must validate the \ac{ACS} certificate against one or multiple root certificates. TR-069 does not discuss which root certificates should be included on a device. In practice, this allows providers to either limit trust to internal \acp{CA} or to rely on the internet public key infrastructure and use a broader public \ac{CA} set, such as Mozilla's \ac{CA} bundle.
	The latter approach can lead to degraded security as all major browsers enforce additional constraints on their trusted root certificates (e.g.\ restricting the certificate issuance date for specific \acp{CA}), which are not reflected in the \ac{CA} file \cite{no_browser_bad_tls}. Additionally, as public \ac{CA} sets change over time, the provider must update the \ac{CA} file. TR-069 does not discuss or mandate any form of certificate revocation.
	
	\item[Protocol Downgrades] TR-069 states that ``if the \acs{ACS} does not support the version with which the \acs{CPE} establishes the connection, it might be necessary to negotiate an earlier \ac{TLS} 1.x version, or even \ac{SSL} 3.0'' \cite[p. 41]{tr069}. While the specification states that \ac{TLS} 1.2 \emph{should} be used, it is very possible that some clients will in practice be susceptible to downgrade attacks to \ac{SSL} 3.0.
	
	\item[Client Certificates] The use of client certificates to identify a client is optional. Client certificates can be unique per device or shared between devices. In the latter case, the specification \emph{recommends} -- and does not require -- to additionally authenticate the client using \ac{HTTP} basic or digest authentication.
\end{description}

\noindent
Notwithstanding these issues, TR-069's major weakness with regard to connection confidentiality and integrity is that the use of \ac{TLS} is optional. This is particularly severe as the \ac{ACS} \ac{URL} can often be reconfigured by various means (see Section~\ref{sec:acs-discovery}), which allows an attacker to downgrade the protocol to plaintext. If \ac{TLS} is used, the use of nonstandard certificate verification methods, missing guidance on root \ac{CA} selection, proposed \ac{SSL} 3.0 support, and shared client certificates may also degrade the security of the connection to a point where man-in-the-middle attacks are feasible.

\paragraph{Authentication and Initial Inform}
\label{sec:preliminaries-tr069auth}
By definition, the first message of every TR-069 session is an \texttt{Inform} command sent by the client (see Appendix \ref{lst:tr069example} for an example). The initial message identifies the device, lists the events that caused the connection, and informs the \ac{ACS} about configuration parameters, such as the current public \ac{IP} address. 

If the client has not used a client certificate to authenticate itself during the \ac{TLS} handshake, TR-069 mandates that the \ac{ACS} must require \ac{HTTP} authentication. The specification \emph{recommends} to use a combination of organization identifier and device serial number as the username and states that clients \emph{should} use unique per-device passwords. Clients \emph{must} use \ac{HTTP} digest authentication for plain \ac{HTTP} connections, but also \emph{must} support \ac{HTTP} basic authentication, which is \emph{recommended} if the connection is secured using \ac{TLS}. If clients erroneously handle basic authentication challenges over unencrypted \ac{HTTP}, %
a man-in-the-middle attacker can exfiltrate the client's username and password. We discuss the security of shared credentials in Section \ref{sec:inspect-emulated-client}.

\paragraph{Commands} After the initial \texttt{Inform}, TR-069 defines further commands that can be used for the communication between client and \ac{ACS}. At the minimum, a conforming \ac{ACS} must support the \sloppy{\texttt{Inform}, \texttt{Get\-\acs{RPC}\-Methods} and \texttt{Transfer\-Complete} commands. 
For clients, the requirements are more extensive. 
A client must support
	\texttt{Get\-\acs{RPC}\-Methods}, \texttt{Get\-Parameter\-Names}, \texttt{Get-/Set\-Parameter\-Values}, \texttt{Get-/Set\-Parameter\-Attributes}, \texttt{Add-/Delete\-Object}, \texttt{Reboot}, and \texttt{Download}}. 
Additionally, TR-069 defines optional commands as well as the ability to specify vendor extensions. 
We describe commands that are interesting from a security and privacy perspective in \CommandTable.

Using the \texttt{Download} command, an attacker has complete control over devices as new firmware can be uploaded that circumvents any potential access restrictions. Even without a special firmware, TR-069 grants network providers access to valuable data on their customers, ranging from detailed information on devices connected to a router to television viewing statistics. These capabilities for data collection are not an inadvertent byproduct of generic commands. Collection of audience statistics is an explicit part of TR-135, which specifies the TR-069 data model for set-top boxes \cite{tr135}. Given the sensitivity of the data at hand, it is reasonable to take a look at which parameters providers regularly access. In the next section, we will discuss how TR-069 communication can be intercepted and inspected.

\section{TR-069 Inspection}
\label{sec:inspect}

Many TR-069 devices are embedded systems which do not provide immediate means to analyze their inner workings. Our first contribution to research is a systematic exploration of three methods that can be used to intercept and monitor TR-069 traffic: man-in-the-middle attacks, client reconfiguration, and client emulation.

\subsection{Man-In-The-Middle Attacks}
\label{sec:inspect-mitm}

When locked-down devices do not provide any means for inspection, researchers can resort to the analysis of network traffic. 
The analyst controls a machine between the TR-069 client and the \ac{ACS}, allowing him to observe all communication. For unencrypted TR-069 sessions, this setup is sufficient to read and modify all commands and their responses. If the connection is protected with \ac{TLS}, an analyst can mount a man-in-the-middle attack against the client to achieve the same end, e.g.\ by using mitmproxy~\cite{mitmproxy}. While this method relies on the availability of trusted certificate keys or the presence of security vulnerabilities on the client, previous research has shown that this is often the case for non-browser \ac{TLS} implementations (see Section~\ref{sec:related-work}). For TR-069 clients in particular, we have outlined multiple potential weaknesses in Section~\ref{sec:preliminaries-security}.

Next, we study the use of \ac{TLS} by service providers before we discuss the applicability of traffic interception on the physical layer for a specific class of TR-069 devices, namely internet gateways.

\begin{figure}[p]
	\caption{%
		TR-069 Provider Configurations\label{fig:providerconfs}%
	}
	\newcommand{\providerFalse}{}
	\newcommand{\providerTrue}{•}
	{
		\scriptsize
		\centering
		\def\arraystretch{0.9}
		\setlength{\tabcolsep}{.25em}
		\begin{tabular}{llccccccccccc}
			\toprule
			\multicolumn{2}{l}{} & \multicolumn{5}{c}{Configuration} & \multicolumn{3}{c}{Firmware} & \multicolumn{3}{c}{Verification}\\\cmidrule(rl){3-7}\cmidrule(rl){8-10}\cmidrule(rl){11-13}
			Country & Provider & NOHC & TLS & VER & PIN & CRT & 2016 & 2017 & 2018 & PUB & TOR & MAIL  \\
			\midrule
			\input{resources/acs_list_generated.tex}
			\\\midrule
			\# Total & 60 & 22 & 31 & 16 & 2 & 2 & 37 & 50 & 60 & 11 & 4 & 60
			\\\midrule
			\multicolumn{13}{l}{
				\begin{minipage}{12.65cm}
					\textbf{NOHC:} Firmware does not contain hardcoded credentials;
					\textbf{TLS:} Client should use \ac{TLS};
					\textbf{VER:} Client should verify server certificate;
					\textbf{PIN:} The number of trusted root certificates is reduced;
					\textbf{CRT:} Use of a (shared) client certificate;
					\textbf{2016-2018:} Configuration has been found in firmware images released in the respective year;
					\textbf{PUB:} The \ac{ACS} is reachable from the public internet;
					\textbf{TOR:} The \ac{ACS} is reachable from selected Tor exit nodes;
					\textbf{MAIL:} Provider has been informed of results and did not rebut findings.
				\end{minipage}
			}
			\\\bottomrule
		\end{tabular}
	}
\end{figure}

\subsubsection{Use of \acs{TLS} by Service Providers} 
\label{sec:inspect-mitm-tlsuse}
Recall from Section~\ref{sec:preliminaries-security} that \ac{TLS} is optional for TR-069. Hence, it is interesting to examine how many providers  follow the recommendation to use \ac{TLS} \cite[p. 40]{tr069} in practice. In search of a data source, we observed that several aftermarket routers sold in Europe\footnote{Germany, for example, passed a law establishing freedom of choice for routers in November 2015 \cite{routerchoice}. Therefore, Germany has a strong ecosystem of aftermarket routers which are user-friendly to set up. The situation in other European countries is similar.} prompt users for their provider's name during setup and use this information to configure basic connection parameters of the device, including its TR-069 settings. 
These settings are also contained in the firmware update packages that many router manufacturers make available on the internet.
We systematically downloaded current and outdated firmware update packages released between 2016 and 2018 for various \ac{DSL}, cable, and fiber modems as well as routers, extracted the file systems, and searched them for TR-069 credentials. However, as there is substantial overlap between the providers supported by aftermarket routers, we are confident that we have a comprehensive coverage of providers in the relevant markets.

The results of this analysis are shown in \AcsTable. Overall, we obtained 471 TR-069 configurations representing 60 internet service providers. Our sample set likely over-represents European countries (Germany, in particular) as well as larger service providers that are more likely to be included in a router's setup assistant. 
Strikingly, we found that almost half of the providers do not use \ac{TLS} at all (29 of 60). Among those who use \ac{TLS}, almost half do not validate server certificates (15 of 31).
Two providers limit the number of trusted \acp{CA} to their own authority, whereas all other providers trust the default set of \acp{CA} installed on the respective device. Two other providers mandate the use of \ac{TLS} client certificates. However, they use a shared certificate for all devices, the private key of which can be extracted from the firmware in both cases. While larger providers in our data set seem to use \ac{TLS} more prudently, we observe that providers often do not adequately protect TR-069 sessions against man-in-the-middle attacks.

We used multiple methods to verify the validity of the provider configurations we found. First, we attempted to establish a \ac{TCP} connection with all \acp{ACS} in our data set to verify their presence. This confirmed the existence of only 11 configuration servers, as many providers presumably have firewall rules in place to deny access from foreign networks. Second, we used the Tor anonymity network to scan for \acp{ACS} from within other providers' networks. Even though the vast majority of Tor exit nodes is not located in residential \ac{IP} space, this confirmed four additional servers. We presume that we would be able to confirm substantially more \acp{ACS} by using ``residential proxy services''. We did not use these proxies as they are -- very similar to DDoS booter services -- universally run by criminals \cite{thomas2017ethical}. %
Finally, we informed each provider of our findings and provided them with the opportunity to correct invalid or outdated results in a responsible disclosure process (see Appendix~\ref{sec:ethics}). 

Another option to find additional \acp{ACS} would have been a scan of the entire IPv4 address space for public instances. As port 7547 is used by both TR-069 clients and configuration servers, we would have needed to interact with these systems to determine their nature. We have refrained from doing this because of the unacceptably high risk of interfering with fragile TR-069 systems in the wild.
For example, Weinmann's analysis of an alleged TR-064/069 vulnerability in 900k routers demonstrated that just probing port 7547 in a specific way can result in an effective DoS attack against routers and large-scale outages \cite{weinmann}.

\subsubsection{Man-In-The-Middle Attacks on Internet Gateways}
A major share of all TR-069 clients are internet gateways. 
Launching a man-in-the-middle attack against these devices on customer premises can be tricky 
because it requires -- depending on the device type -- the interception of copper telephone lines, coaxial cable, or optical fiber. In contrast to Ethernet connections, which are easily intercepted with off-the-shelf hardware, sniffing is not straightforward in these cases. Cable internet, for example, operates on a shared medium, therefore providers often encrypt the traffic between the cable modem and their cable termination system. Although previous research has shown that cable encryption is often weak \cite{security_of_cable_networks_case_study,cable_modems_later_years}, without doubt such attacks remain significantly more complex than with Ethernet; not to mention the ethical and legal concerns related to eavesdropping on a shared medium. As a workaround, gateways with integrated routers can often be configured to disable the modem and treat a regular \ac{LAN} port as their (internet) \ac{WAN} port instead. In this case, the router can be moved into an existing network where it can be monitored. Nonetheless, a ``classical'' interception of internet gateways remains difficult to implement in general. An oftentimes easier approach is the use of client reconfiguration, which we will discuss in the next section.

\subsection{Client Reconfiguration}
\label{sec:inspect-clientreconf}

\noindent
When analyzing router firmwares for TR-069 configurations, we found that many devices provide means to modify their TR-069 settings. This makes it possible to perform a considerably easier man-in-the-middle attack. Instead of intercepting the connection between gateway and \ac{ACS}, we place a machine in the local network that acts as a reverse proxy to the original \ac{ACS}. We then reconfigure the gateway to use this machine as its \ac{ACS}. In other words, we instruct the gateway to use \texttt{http://acs.example.local/} as its \ac{ACS}, where \texttt{acs.example.local} is our local proxy that forwards all requests to the provider's \ac{ACS}. This has a number of advantages:
\begin{enumerate}
	\item For modems, this removes the requirement to perform a man-in-the-middle attack on copper telephone lines, coaxial cable, or optical fiber. Instead, interception can be done with off-the-shelf Ethernet devices.
	\item By reconfiguring the \ac{ACS} \ac{URL}, one can downgrade the protocol between gateway and interception device to plain \ac{HTTP}, which makes it possible to intercept connections otherwise protected by \ac{TLS} (see Section \ref{sec:acs-discovery}).
	\item In contrast to the network setup described in Section \ref{sec:inspect-mitm}, the inspection devices only needs to forward TR-069 communication. As all other network traffic is unaffected, the proxy device needs considerably fewer resources.
\end{enumerate}

\noindent
However, client reconfiguration is not free of pitfalls. The \ac{ACS} may use a TR-069 session to update the server \ac{URL}, which would override the custom redirection and bypass the proxy. In contrast to a classical man-in-the-middle attack, the local proxy device has limited means to detect and correct this. While regular TR-069 \texttt{SetParameterValues} commands can be manipulated on-the-fly, the \ac{ACS} \ac{URL} may also be reset by firmware updates. This makes long-term monitoring of reconfigured devices an ongoing challenge.

TR-069 defines multiple mechanisms to configure the \ac{ACS} \ac{URL} (see Section~\ref{sec:acs-discovery}). Devices differ widely in what they support. For example, on some TP-Link routers, we were able to configure the \ac{ACS} \ac{URL} via the router's web interface. For AVM\footnote{In 2013, AVM's Fritz!Box series had an estimated market share of 68\% in Germany and 18\% in Europe \cite{avm}.} routers, we found that TR-069 can be configured via TR-064 (the configuration interface available from LAN), or by restoring a forged configuration file. In contrast, we did not find  means to reconfigure a Cisco cable modem without intercepting communication on the coaxial cable first. But we were able to reconfigure a Cisco \ac{IP} phone via \ac{DHCP}.

\subsection{Emulated Clients}
\label{sec:inspect-emulated-client}

When observing the traffic of a given TR-069 client is impractical, another option is to completely emulate a TR-069 device. This is particularly useful to repeatedly analyze the handling of selected events such as device provisioning. 

In contrast to the previously described interception techniques, the major challenge with emulated clients is not intercepting connections, but obtaining valid credentials to communicate with the configuration server. In this section, we discuss how clients can authenticate themselves to existing configuration servers to establish TR-069 sessions. We describe our emulated honeyclient later on in Section~\ref{sec:honeyclient}.

To establish a TR-069 session, a client generally needs two pieces of information: the \ac{ACS} \ac{URL} and valid credentials for authentication. The difficulty of obtaining these depends on provider and device. 
If both device manufacturer and provider put a high emphasis on protecting their TR-069 communication against external observers, analysts need to extract credentials by manually reverse-engineering specific devices. 
However, we find that the use of shared credentials is prevalent in practice. When we analyzed the use of \ac{TLS} by service providers in Section \ref{sec:inspect-mitm-tlsuse}, 38 of 60 providers had hardcoded authentication credentials in their TR-069 configuration. Next we discuss how this impacts connection security.

\vspace{-.2cm}
\paragraph{Shared Authentication Credentials}
Counter-intuitively, hardcoded TR-069 credentials do not necessarily imply a vulnerability. If TR-069 is used for monitoring purposes only and not for the provisioning of, for instance, \ac{VoIP} credentials, the secure identification of individual devices is less of a concern. Also, a provider may only use hardcoded credentials to provide an initial generic configuration and then update the credentials stored on the device with unique ones using the \texttt{SetParameterValues} command. The TR-069 specification discourages hardcoded credentials, but does not prohibit them. When sharing credentials between devices, providers need to resort to alternative means of identifying individual customers. There are various attributes a provider can use for this purpose:
\vspace{-.1cm}
\begin{enumerate}[noitemsep]
	\item The client's \ac{IP} address as seen by the server.
	\item The client's \ac{IP} address as reported in the initial \texttt{Inform} command.\footnote{Without taking security concerns into account, a provider may prefer this over option~1 because it works if proxies in the provider's network mask the client's \ac{IP}.}
	\item The device's serial number as reported in the initial \texttt{Inform} command.
	\item The router's \ac{MAC} address, which can be obtained with a \texttt{GetParameterValues} command.
	\item User input collected in a custom authentication dialog.
\end{enumerate}
\vspace{-.1cm}

\noindent
We briefly discuss each option's security properties.
For the first four options, we assume that the provider has reliable means to map \ac{IP} addresses, serial numbers, or \ac{MAC} addresses to customers.

Option 1 has the major issue that any device in a customer's network can obtain the customer's TR-069 configuration. For example, clients in a public wireless network (e.g.\ in caf{\'e}s) could obtain the owner's \ac{VoIP} credentials and commit phone fraud.

Options 2--4 have a common problem in that they rely on information which is easily spoofable and should not be considered secret. The \ac{IP} space of an internet service provider is public knowledge and \ac{MAC} addresses or serial numbers cannot be assumed to be random nor secret. For example, we found that the serial number of an AVM router can be read by any client in its network without authentication.

Option~5 is the most secure one, but complex to implement correctly. The TR-069 specification does not provide any guidance on possible authentication methods. Moreover, many providers are reluctant to use this option as it conflicts with the stated goal of ``Zero-Touch'' configuration (cf.\ Section~\ref{sec:preliminaries-tr069}).

In summary, TR-069 client authentication is a hard problem for service providers in practice.
The use of hardcoded credentials comes with unique security challenges, yet the absence of hardcoded TR-069 credentials does not necessarily imply that the provider is using different or better means to identify customers.
Previous work has indicated that this problem is not of theoretical nature~\cite{o2_acs_auth}. 
The prevalence of hardcoded credentials in practice makes it quite easy to set up emulated clients in many networks.%

\vspace{-.4cm}
\subsection{Comparison of Inspection Methods}
\vspace{-.2cm}
\label{sec:inspect-comparison}

Man-in-the-middle attacks, client reconfiguration, or emulated clients are sufficient to inspect many TR-069 deployments in practice.
However, none of the methods is guaranteed to succeed with a specific client  and configuration server. As a last resort, interception may require manual reverse-engineering of individual devices.
For researchers and practitioners interested in the analysis of TR-069, we make the following recommendations:

\vspace{-.2cm}
\begin{enumerate}
	\item \textbf{For all devices that allow easy network interception,} the most effective approach is to attempt a classic man-in-the-middle attack first. If unsuccessful, reconfiguration via \ac{DHCP} can be tried next. At the very least, this approach provides researchers with the \ac{ACS} address, which can then potentially be matched with TR-069 credentials extracted from other devices.
	\item \textbf{For modems where interception is difficult or man-in-the-middle attacks fail,} we propose to attempt a device reconfiguration via web interface, TR-064, \ac{DHCP}, or settings files (in this order). While our small sample set of devices produced by three different manufacturers does not allow for general conclusions, we were quite successful in downgrading TR-069 to plain \ac{HTTP} and rerouting connections. %
	\item \textbf{For an interactive approach to TR-069 monitoring,} emulated clients can be used to observe how an \ac{ACS} reacts to simulated device events. This is particularly useful to test firmware upgrade behavior, device authentication, and provisioning procedures. The ease of deployment makes emulated clients attractive for large-scale studies.
\end{enumerate}

\noindent
In summary, TR-069 inspection is very feasible in practice, although we are not aware of a ``silver bullet''. Setting up devices for inspection is a highly manual process.

\section{TR-069 Honeyclient}
\label{sec:honeyclient}

While TR-069 inspection by man-in-the-middle attack or client reconfiguration (cf.\ Sections~\ref{sec:inspect-mitm} and \ref{sec:inspect-clientreconf}) can be implemented with standard tools, at the time of writing we are unaware of any existing software that can readily emulate TR-069 clients (cf.\ Section~\ref{sec:inspect-emulated-client}). Here we introduce an open-source honeyclient we developed to solve this problem. Our honeyclient emulates TR-069 devices and can be used to watch configuration servers and %
observe their response to certain events. This allows us to assess how different devices are bootstrapped, which parameters a provider %
accesses, or how fast security updates are rolled out.

The design of the honeyclient is based on three central requirements: full coverage of the TR-069 protocol, usability for the target audience of proficient security analysts, and the capability to perform deliberate protocol violations. We discuss the implementation of these requirements in Appendix \ref{sec:honeyclientimpl}. 
For an initial exploration, we provide an interactive command line interface that can be used to issue TR-069 commands and observe the provider's response. Alternatively, the honeyclient can be instrumented in Python.
Next, we describe the results of using our honeyclient for the automated security testing.

\subsection{Analysis of Open-Source TR-069 Servers}
To demonstrate that our client is effective at uncovering security vulnerabilities, we have used its scripting capabilities to develop an automated test suite for configuration servers, which we instantiated in a controlled network environment. Our targets are GenieACS \cite{genieacs}, the only open-source configuration server in active development, and OpenACS \cite{openacs}. The latter is  still relevant as we found multiple providers with an \ac{ACS} \ac{URL} indicating that they likely run OpenACS.

Even though we only performed very limited automated black-box tests with our honeyclient, we found security vulnerabilities in both tested open-source configuration servers. %
Our findings are described  in Appendix \ref{sec:honeyclient-serveranalysis}.
While we note that the vulnerabilities are specific to GenieACS and OpenACS, empirical research has shown that open-source and closed-source software often do not significantly differ with regard to the severity of vulnerabilities \cite{anderson2005open,schryen_foss_security_myth}.
We suspect that the use of serial numbers for customer identification might be a widespread issue with configuration servers as the TR-069 specification does not provide any guidance on client authentication.
This is critical as it allows attackers to e.g.\ obtain other user's \ac{VoIP} credentials for phone fraud \cite{phonefraud}.

\section{TR-069 Infrastructure Monitoring}
\label{sec:monitoring}

When looking at TR-069 from a security and privacy standpoint, it is interesting to not only analyze a single client, \ac{ACS} implementation, or provider in isolation, but to compare how different providers and clients behave over time and in reaction to specific events. For example, how fast are different providers rolling our security updates? How many providers use TR-069 not only for remote management, but also to obtain information on sensitive data such as television viewing statistics? How is TR-069 used in countries with restricted freedoms to implement censorship or surveillance?
Monitoring TR-069 infrastructure 
at a larger scale
provides better information on the security and privacy aspects of TR-069 in practice.
In this section, we propose how our toolkit can be turned in an infrastructure for monitoring a larger number of TR-069 deployments over longer periods of time. For brevity, we describe the implementation of our individual monitoring sensors based on inexpensive mini routers in Appendix \ref{sec:monitor-sensor} and our virtual test environment that makes TR-069 more accessible to researchers in Appendix~\ref{sec:virtual-test-env}. Based on our monitoring sensors, we outline the challenges of large-scale TR-069 monitoring and report measurements of twelve months of real-world TR-069 traffic. %
\subsection{Distributed Monitoring System}
\begin{figure}[t]
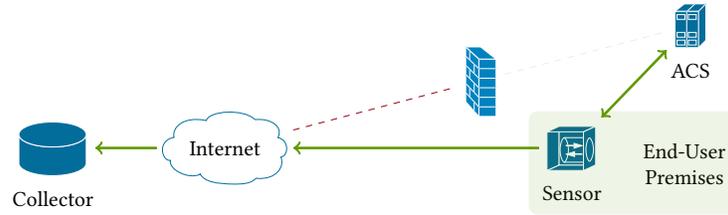

	\centering
	\includestandalone[page=6,width=.8\linewidth]{resources/illustrations}
	\caption{
		\label{fig:monitoring-architecture}
		Data collection setup
	}
\end{figure}

\noindent
Compared to internet monitoring efforts such as Censys \cite{censys15}, collecting data on the TR-069 infrastructure is more challenging because many configuration servers are not reachable from the public internet. To support this with data:
among server addresses we obtained in the analysis of Section~\ref{sec:inspect-mitm-tlsuse}, we found that 49 out of 60 hosts were shielded off the internet, presumably to reduce the attack surface of the \ac{ACS} (see \AcsTable). Therefore, a monitoring infrastructure must be distributed with sensors located in each provider's network. 
Additionally, the installation of sensors requires some expertise and manual configuration: for each deployment, either a suitable inspection method or valid authentication credentials for a honeyclient need to be obtained. For TR-069, this often involves reverse-engineering individual devices and thus limits the scalability of our approach to TR-069 monitoring. However, we believe that a limited network of sensors is still useful to observe basic TR-069 operations in practice. %

To collect data from multiple sensors, we have opted for a simple centralized logging infrastructure. Sensors report all TR-069 activity to a dedicated log collector via \ac{HTTPS} (see Figure \ref{fig:monitoring-architecture}).
Data is redacted before transmission (cf. Appendix~\ref{sec:ethics}) and then stored on the collector.

\subsection{Real-World Measurement Study}\label{sec:measurement-study}

To evaluate our monitoring system, we have conducted a twelve month study where we continuously monitored TR-069 traffic of ten households in Western Europe. Study participants were recruited through the first authors' personal contacts and informed what kind of data will be collected. For each participant, we determined a suitable inspection method for the respective router and configured a monitoring sensor on-site. In total, our measurements cover nine distinct device models and five internet service providers (cable and DSL) starting in March 2018. Our main findings are as follows:

Over the course of our study, three device models received updates from the manufacturer. However, we did not observe any provider deploying these updates to our participants' devices via TR-069. Two devices were manually patched by their owners four and twenty days after firmware release respectively, and one device remains unpatched for ten months as of today (December 2019). A fourth device entered our study in an unpatched state and still remains so. While TR-069 could improve end user security by providing timely firmware updates, we found that providers are not doing their homework as we couldn't observe any evidence of this in practice.

In more positive news, we did not find any ``smoking gun'' indicating severe privacy violations by service providers. Next to the periodic \texttt{Inform} messages sent by clients, providers sent commands that can all be reasonably linked to regular maintenance work. For example, two providers updated the \ac{ACS} URL to a new configuration server while a third provider repeatedly queried DSL connection quality parameters for a period of two days to presumably debug connectivity issues. While our findings show no evidence of widespread privacy violations, providers could still target particularly interesting customers individually.
Overall, our study validates that the distributed monitoring system works reliably for long-term TR-069 monitoring.

\section{Discussion}
\label{sec:discussion}

As the first contribution of this paper, we have discussed man-in-the-middle attacks, client reconfiguration and emulated clients as three methods to inspect TR-069 traffic. 
Our analysis of provider configurations has shown that providers often neglect to follow the security recommendations made in the TR-069 specification. We hypothesize that this is happening because plain \ac{HTTP} is still marginally cheaper and simpler to implement while sniffing or man-in-the-middle attacks are not perceived as realistic threats.
All devices we analyzed would have supported \ac{TLS}. From our point of view, allowing plain \ac{HTTP} is a design mistake of the TR-069 specification which opens the door to a whole class of attacks.  %

We do want to point out that -- notwithstanding our previous point -- interception and monitoring of internet gateways can still be difficult to establish when devices are resistant to reconfiguration. There are two reasons why we believe that this is not as much of an issue as one may think: First, TR-069 has moved from being used by internet gateways only to supporting a whole range of devices such as set-top boxes and \ac{VoIP} phones, which all can be intercepted more easily on Ethernet. Second, the use of emulated clients often is an effective alternative to monitoring real devices.

The second contribution of this paper is an open-source TR-069 honeyclient that can be used to assess TR-069 configuration servers from a security and privacy point of view. As analysts can emulate arbitrary events, we found that it is often more effective to use a honeyclient when analyzing a configuration server than to monitor a real device. We further believe that our client is not only effective at monitoring, but also at offensive security research, as evidenced by the vulnerabilities we found in OpenACS and GenieACS using automated black-box testing. The security of configuration servers is a promising avenue for future research.

\noindent
With regard to practical TR-069 security, our most important finding is the widespread use of hardcoded credentials. Both GenieACS and OpenACS use the device serial number as the primary means of identification, and the prevalence of hardcoded credentials in provider configurations suggests that this is also the case for a variety of commercial configuration servers.
As we discussed in Section \ref{sec:inspect-emulated-client}, neither the use of \ac{IP} address-based nor serial number-based authentication can be considered secure.
At the same time, the TR-069 specification does not provide any guidance on how clients should be identified. 
When clients are not authenticated securely, an attacker can impersonate other users to obtain their TR-069 configuration, which includes for example \ac{VoIP} credentials.

As the third contribution of this paper, we have developed a monitoring sensor and a centralized log collector to monitor TR-069 deployments in practice. We have demonstrated that our sensor can be run on inexpensive routers, but we did not perform a rigorous evaluation of possible hardware alternatives. As our sensor runs on OpenWrt, we believe that it can be easily adapted to work on other hardware platforms. %
A major concern in the development was the privacy of study participants (see Appendix \ref{sec:ethics}). The sensitive nature of TR-069 traffic makes it difficult to conduct analyses that are both comprehensive and privacy-preserving. At the very least, the inner workings of our platform are -- down to the operating system -- open-source software and available for inspection. 

An interesting avenue for future work is TR-069 monitoring at high-value endpoints that may be targeted with custom configurations. Here we see two main challenges that need to be addressed in future research. First, our sensor deployment currently has high setup costs. Sensors must be installed on customers' premises, installation is a relatively manual process, and user privacy must be considered appropriately. Second, when asking volunteers to set up sensors and contribute data, the authenticity of our results is at stake as we cannot verify the legitimacy of logs with our current architecture. Nonetheless, we believe that even relatively small sensor deployments can yield interesting insights in practice (see Section~\ref{sec:measurement-study}). With the growing adoption of TR-069 for devices other than internet gateways, monitoring TR-069 becomes gradually more important and easier to accomplish at the same time.

\newpage
\section{Related Work}
\label{sec:related-work}
\vspace{-.1cm}

Despite the widespread use of TR-069, there is very little public information about the protocol's security and privacy implications. In this section, we first review works on TR-069 security and then discuss publications on man-in-the-middle attacks and honeypots that inspired our work. 

\vspace{-.1cm}
\paragraph{TR-069 Security}  We are not aware of any academic publications on TR-069 security. %
Therefore, we deem it also valuable to point to ``grey'' literature.

One team of security researchers has focused its attention on specific implementation vulnerabilities in TR-069 clients and configuration servers. Most notably, Tal presented a talk on exploiting TR-069 configuration servers at Defcon~22 \cite{defcon22shahar} in 2014. He disclosed vulnerabilities in multiple configuration servers and gained control over 500,000 managed devices in one particular instance. The respective security advisories are still undisclosed as of December 2019. In a follow-up talk, %
Tal and Oppenheim demonstrated the ``Misfortune Cookie'' vulnerability in the RomPager~TR-069 connection request server, which is embedded in at least 12 million residential gateways manufactured by ASUS, D-Link, Edimax, Huawei, TP-Link, ZTE, ZyXEL, and others they found exposed to the internet \cite{31c3cooks}. 
While RomPager developer AllegroSoft issued a version that fixes the vulnerability in 2005, devices today (2019) still ship with the vulnerable version. This shows that the patch propagation cycle is incredibly slow (sometimes non-existent) for these types of devices \cite{misfortunecookie}. %

In January 2016, RedTeam Pentesting published an information disclosure vulnerability in o2/Telefonica Germany's configuration server \cite{o2_acs_auth}. QA Cafe, a US-based software company specialized in testing broadband gateways, publishes TR-069 security best practices for service providers \cite{qacafebestpractices}. We also note \"{O}mer Erdem's master thesis on HoneyThing, a honeypot that emulates three known vulnerabilities in connection request servers \cite{honeything}. Vetterl and Clayton use firmware images to emulate CPE/IoT devices and run them as honeypots \cite{vetterl2019honware}.

\vspace{-.1cm}
\paragraph{Man-In-The-Middle Attacks on \ac{TLS}} 
Man-in-the-middle attacks are a common technique to learn or modify communication contents that are otherwise protected using \acf{TLS} \cite{mitmbasics}. %
Previous research has shown that non-browser \ac{TLS} implementations often implement certificate verification incorrectly and are subject to such attacks \cite{fahl2012androidevemallory,most_dangerous_code}. For a comprehensive overview of past \ac{TLS} security issues, we refer to Clark and Van Oorschot \cite{sok}.

\vspace{-.1cm}
\paragraph{Honeyclients} A honeypot is ``a security
resource whose value lies in being probed, attacked,
or compromised'' \cite{lance}. One subgroup of honeypots are honeyclients, which mimic the behavior of a network client that actively initiates requests to servers aiming to be attacked \cite{seifert2006taxonomy}. The development of our TR-069 honeyclient was largely inspired by Nazario's PhoneyC, a honeyclient that emulates a web browser to enable the study of malicious websites \cite{phoneyc}. Similar to PhoneyC, our honeyclient emulates a TR-069 device and connects to configuration servers to get attacked. In contrast to most existing honeypots, the value of our honeyclient lies not only in the detection of  attacks, but also in the logging of events that demonstrate privacy infringements or other unwanted actions by providers.

\newpage
\section{Conclusion}
\label{sec:conclusion}
This paper takes a look at the security and privacy aspects of the TR-069 remote management protocol. Although the protocol is used in nearly a billion devices, our work represents (to the best of our knowledge) the first academic publication on the topic. 

To facilitate further research, we have first discussed the protocol from a security perspective and presented three methods researchers can use to analyze the TR-069 communication of their devices.
None of the approaches is guaranteed to succeed with a specific client or configuration server, but we have shown that man-in-the-middle attacks, client reconfiguration and the use of emulated clients can be used to inspect the majority of networks in practice. To enable analyses that go beyond the monitoring of existing devices, we contribute an open-source honeyclient implementation that can be used to assess configuration servers. Using our client, we were able to obtain information on firmware updates, \ac{VoIP} credentials, and data access patterns by providers in practice.

While its primary field of use lies in the monitoring of configuration servers, our honeyclient can also be utilized by providers to debug and stress-test their configuration servers, as well as researchers who can instrument it for offensive security research. 
We have built an automated test suite on top of our client to demonstrate weaknesses in OpenACS and GenieACS, two open-source configuration servers. 
The vulnerabilities we found provide initial evidence that client authentication could be a systemic issue in the TR-069 landscape.
Our findings support the general assertion by Tal and Oppenheim that TR-069 infrastructure is most often inadequately secured \cite{31c3cooks}. 

To facilitate the monitoring of providers for privacy infringements and other unwanted behavior conducted over TR-069, we have developed a monitoring sensor based on mitmproxy.
Compared to internet-wide scanning studies, the manual work required to deploy individual sensors is a fundamental limitation for large-scale TR-069 measurements. We release both our honeyclient and the sensor software as open source under the MIT license\footnote{\url{https://github.com/mhils/tr069}}.

We have used our monitoring system to conduct real-world measurements of TR-069 traffic for twelve months. While we did not observe any privacy violations by providers, we also found no evidence of providers using TR-069 to push firmware updates released by vendors during our measurement period. The purported security benefits of TR-069 are not realized.

Looking ahead, we hope that our work lays the foundation for more security research on TR-069. We limited our real-world measurements to routers as the most popular TR-069 device type, but we expect research on for example set-top boxes to yield interesting data on potential privacy infringements by providers. Furthermore, we encourage researchers to use our honeyclient for the assessment of TR-069 configuration servers. As these systems effectively have remote execution privileges on all connected devices, they represent high-value targets that would benefit from more throughout security analyses in practice.

\newpage 
\appendix

\section{Ethical Considerations and Privacy Measures}
\label{sec:ethics}
For the research of this paper, we collected providers' TR-069 credentials and deployed sensors to monitor our study subjects' TR-069 communication. This makes it inherently challenging yet critically important to execute our research ethically. While our host institution has processes in place, they do not apply to our research. Researchers are advised to act responsibly in in compliance with data protection laws; we chose to follow the Menlo Report \cite{menlo} and, when necessary, consulted legal experts.

\paragraph{TR-069 Provider Configurations} While we did not test providers' ACS deployments for specific security vulnerabilites, we have notified all providers for which we obtained TR-069 configurations (see Section \ref{sec:inspect-mitm-tlsuse}) about our findings as the publication of our results may point criminals to potential opportunities for abuse. Providers were made aware of our upcoming publication and were provided with an anonymized version of \AcsTable{} in which their own identity was revealed to them. As many providers did not provide security contact email addresses, all notifications were sent by postal mail and providers were given 60 days to comment. We received positive replies from multiple providers acknowledging that this is an industry-wide issue. No providers objected to publication or debated our results. We believe that publishing data on the current security level of the TR-069 ecosystem is necessary to incentivize improvements. While useful for future research, we refrain from publishing any of the collected \ac{ACS} credentials as they could be easily abused.

\paragraph{Real-World Measurement Study}
A key challenge with our measurement study was to retain user privacy while collecting relevant data. To protect study participants' data, we have implemented the following measures. All study participants were specifically selected to have a background in Computer Science. They were then informed on how TR-069 works, and what kind of data we were about to collect. They were allowed to withdraw from our study at any point. At the beginning of the study, participants were provided with a sensor device which recorded all TR-069 commands locally and only sent the command names and, in the case of \texttt{Inform} commands, the causing event type, to the collector. For the locally recorded logs, any detected credentials were replaced with placeholders before data was written to disk.

By default, the authors had no access to the sensor devices after permanently handing them over to study participants. If command names pointed to interesting behavior, study participants were asked to inspect the data in the web interface (which is why we required a technical background) and, if they were comfortable sharing it, flip a hardware switch on the sensor device to temporarily start an SSH server the authors could connect to. By using a hardware switch and an indicator light, study participants were at all times able to control access to the sensor and their data. At the end of our study, participants were provided with instructions on how they can delete all collected data and repurpose the devices.

\section{Honeyclient Implementation}
\label{sec:honeyclientimpl}

\begin{figure}[h!]
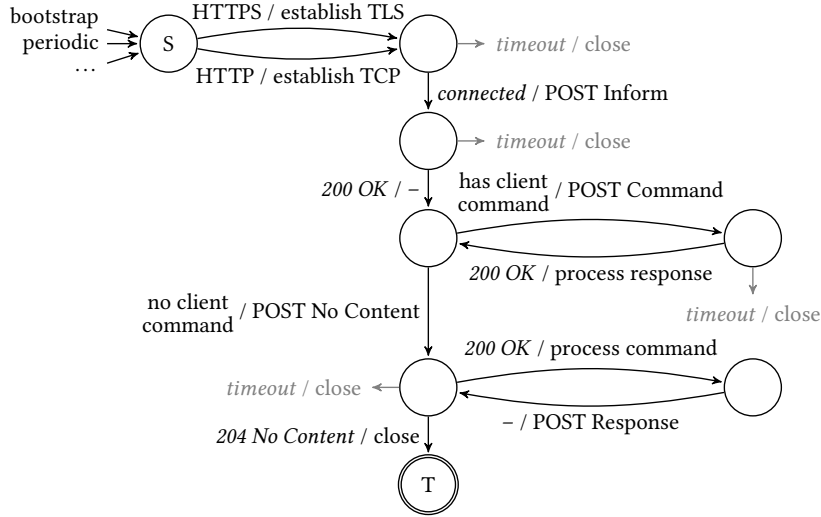

	\vspace{-.8cm}
	\centering
	\includestandalone[page=8,width=.9\linewidth]{resources/illustrations}
	\caption{
		\label{fig:honeyclient-states}
		State machine of conforming TR-069 sessions.
	}
	\vspace{-.4cm}
\end{figure}

\noindent
The design of the honeyclient is based on three central requirements: full coverage of the TR-069 protocol, usability for the target audience of proficient security analysts, and the capability to perform deliberate protocol violations. In the following, we describe how these criteria are reflected in our implementation.

\paragraph{Full Protocol Coverage}
To communicate with all types of configuration servers, our implementation supports all authentication methods described in the TR-069 specification (\ac{HTTP} basic authentication, \ac{HTTP} digest authentication, and \ac{TLS} client certificates), and allows to modify all relevant attributes of the initial \texttt{Inform} message (see Section~\ref{sec:inspect-emulated-client}). Our honeyclient can emulate arbitrary devices, whose configuration can be automatically parsed from intercepted network traffic. It implements every command we have observed in practice. In addition, users can process custom messages, such as vendor extensions. In order to support provider-initiated TR-069 sessions, the honeyclient includes a server component that handles incoming connection requests. Note that the full protocol coverage helps to prevent simple fingerprinting, but stealthiness was not a design goal. Therefore, we cannot fully rule out that an \ac{ACS} can detect that it is talking to our honeyclient instead of the expected device, e.g., by measuring client response times.

\paragraph{Usability}
To facilitate uptake by other researchers, usability is an explicit design goal of our honeyclient. %
This influences our implementation as follows: first, we have opted to write the honeyclient in Python 3, a high-level programming language many security practitioners are familiar with. The Python package has minimal dependencies and can easily be deployed on inexpensive embedded devices. Second, we provide an interactive command line interface that can be used to get familiar with our honeyclient. %
For an initial exploration, researchers can issue TR-069 commands and observe the provider's response, as well as react to out-of-band connection requests. For automated (long-term) monitoring of configuration servers, the client can then be instrumented using Python scripts. To make the analysis of configuration servers more efficient, the honeyclient can automatically process all server commands we have observed in practice. %

\paragraph{Protocol Violations}
Conforming TR-069 clients exactly implement the state machine for TR-069 sessions as shown in Figure \ref{fig:honeyclient-states}. %
The honeyclient can be used to deliberately violate a configuration server's expectations of state transitions. For example, the client can send conflicting commands over the same connection, skip state transitions (e.g.\ the initial \texttt{Inform} message), or replay a previously sent message. This makes it possible to uncover logic errors, missing replay protection, and other issues in the configuration server. It is also possible to craft malformed \ac{SOAP} messages, either to mirror bugs in actual devices (of which we encountered several), or probe systems for \ac{XML} parsing vulnerabilities.

\section{Analysis of Open-Source TR-069 Servers}
Using only automated black-box tests, our honeyclient detected the following vulnerabilities in GenieACS and OpenACS:\footnote{The authors are committed to responsible disclosure. The details of potential weaknesses and discovered vulnerabilities were shared with the maintainers, providers, and the relevant CERTs.}
\label{sec:honeyclient-serveranalysis}

\begin{enumerate}
	\item GenieACS does not implement \ac{HTTP} authentication for TR-069 sessions. Providers running GenieACS must rely on alternative means to identify users, which are often not implemented securely (see Section~\ref{sec:inspect-emulated-client}). 
	\item GenieACS reveals the device configurations of other users when presented with a spoofed serial number in the initial \texttt{Inform} command. An attacker can access arbitrary customer configurations (including e.g.\ \ac{VoIP} credentials) by iterating over candidate serial numbers.
	\item GenieACS is vulnerable to XML External Entity (XXE) attacks, enabling local file disclosure.
	\item GenieACS permits any client command after an initial Inform. Our honeyclient could successfully issue different \texttt{Inform} commands in a single session, allowing us to confuse the server-side state belonging to different devices.
	\item When issuing connection requests, GenieACS accepts a \ac{HTTP} basic authentication challenge from our honeyclient. This violates the TR-069 specification, which mandates digest authentication (see Section~\ref{sec:protocol-connection-establishment}). As basic authentication provides no replay protection, devices and providers are susceptible to denial-of-service attacks. %
	\item OpenACS does not invalidate digest authentication nonces, which permits replay attacks.
	\item Like GenieACS, OpenACS accepts any command at any point during a session, making it susceptible to attacks involving state confusion.
\end{enumerate}

\section{Sensor Implementation}
\label{sec:monitor-sensor}

\noindent
We have developed a monitoring sensor that can be used to intercept and record TR-069 sessions. The sensor records all \ac{HTTP} messages and connection metadata (e.g.\ server certificates) in an append-only transaction log, which can be displayed to the user and exported for analysis. On a technical level, our sensor is implemented as a series of independent add-ons for mitmproxy \cite{mitmproxy}, an open-source man-in-the-middle proxy. In addition to the add-ons used for data processing, we have modified mitmproxy's web interface to better visualize TR-069 sessions.%

The overarching requirement for our sensor design was to facilitate deployment on a wide user base. From this we derive three design criteria: independence of the inspection method, privacy, and inexpensive sensor hardware. In the following, we describe how these criteria are reflected in our implementation.

\paragraph{Independence of Inspection Method}

It is often attractive to combine the inspection methods introduced in Section~\ref{sec:inspect} in a single analysis. For example, when using honeyclients for their ease of set up, it is useful to validate findings by comparing the traffic to traffic generated by real devices.
Our sensor can perform transparent man-in-the-middle attacks, act as a reverse proxy for reconfigured clients, or operate with emulated devices, such as our honeyclient. %
When the sensor is configured as a reverse proxy, it automatically rewrites commands that read or write reconfigured parameters. For example, when a \texttt{SetParameterValues} command is issued to update the \ac{ACS} \ac{URL} on the client, the sensor will modify the message to retain the device reconfiguration and update its internal forwarding address instead.
While this reconfiguration protection can in principle be sidestepped with firmware updates (see Section \ref{sec:inspect-clientreconf}), the sensor would still record the evasion attempt.

\paragraph{Privacy}
A key challenge with TR-069 sensors is retaining user privacy and security while collecting relevant data. From a security perspective, TR-069 sessions occasionally contain secrets such as \ac{VoIP} credentials, which should not be included in the sensor's logs. %
From a privacy perspective, TR-069 sessions regularly contain personally identifiable information as well as information that may reveals user attitudes and behaviors. We have adopted Langheinrich's \cite{Langheinrich2001} guidelines to handle the challenges in the design phase: %

\begin{enumerate}
	\item \textbf{Notice, Choice, and Consent:} As the immediate purpose of our sensor is data collection, we allege that users understand that they are being sensed when they are physically installed in their homes. However, we do not expect them to initially understand what kind of data is transmitted over TR-069. As such, all data is recorded on the device only and can be reviewed before being shared with researchers. While this view on the data is arguably very technical, we presume that this is in the interest of our (likewise technical) audience. Second, a simple physical switch on the sensor is used for access control so that users can revoke consent to sharing their data at any time. We include this option as previous research has shown that privacy controls are exercised in particular if they are easy to understand \cite{webcamcovering}.
	
	\item \textbf{Anonymity and Pseudonymity:} For many research questions regarding TR-069, it is hard to provide anonymity to study subjects. For example, it is fundamentally difficult to not collect personally identifiable information when assessing the extent of privacy infringements by providers. For large-scale monitoring, we provide users with the option to stay anonymous within a study by reducing message contents to command names, thereby removing all information about the state of the client or connected devices.
	
	\item \textbf{Collection and Use Limitation:} TR-069 configuration servers transmit a variety of credentials for e.g.\ \ac{VoIP} accounts or streaming services. This makes the contents of TR-069 sessions an attractive target for criminals. To protect users, the sensor will replace credentials with placeholder values before logging them to persistent storage. This provides an analyst with a comprehensive view on TR-069 sessions %
	without potentially compromising the security of transmitted account credentials.

\end{enumerate}

\noindent
For researchers who intend to design their own methods for conducting privacy-preserving studies on TR-069, we like to point out that we found the reliable redaction of sensitive message contents to be unexpectedly difficult. Two devices we assessed produced \ac{XML} that was malformed to the point where we were unable to parse it consistently with multiple \ac{XML} parsers, e.g.\ because of unescaped control characters or attribute definitions in closing tags.

\paragraph{Sensor Hardware} 
For monitoring TR-069 deployments over longer periods of time, it is desirable to run the sensor on inexpensive, always-on devices. As TR-069 operates 24/7, using end users' computers as a sensing platform is not practical. To demonstrate that our sensor can be run on small embedded devices, we initially experimented with the Raspberry Pi 3, a single-board ARM computer. However, when used in a man-in-the-middle setup, where non-TR-069 traffic needs to be forwarded, we found its network performance unsatisfactory, supposedly because its second \ac{LAN} port is interfaced over \ac{USB}. We ported our sensor to OpenWrt \cite{openwrt}, which can be run as operating system on a variety of consumer and professional wireless routers. %
In contrast to single-board computers targeted to hobbyists, we found that -- perhaps not unexpectedly -- routers are considerably better suited for our man-in-the-middle scenario as traffic forwarding can be offloaded to the network interface. 
\begin{figure*}[htb!]
	\centering
	\subfloat[Plaintext \ac{HTTP} sessions]{{\includegraphics[page=1,width=0.45\linewidth]{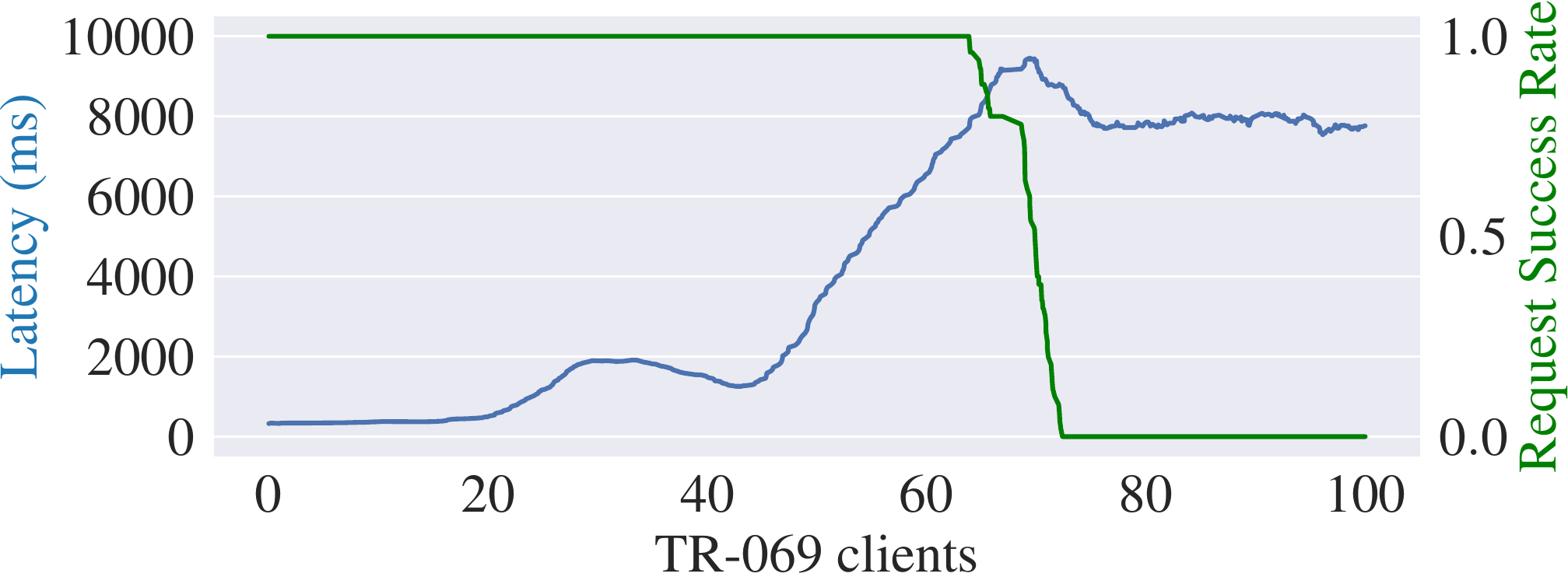} }}%
	\qquad
	\subfloat[\ac{TLS}-protected sessions]{{\includegraphics[page=2,width=0.45\linewidth]{resources/benchmark.pdf} }}%
	\caption{
		Evaluation of sensor performance: man-in-the-middle TR-069 monitoring with 90\,Mbit/s load of forward traffic. 
		\label{fig:sensorperf}
	}%
\end{figure*}

We have used GL Tech's GL-AR300M mini router to test the feasibility of embedded sensor deployments. It runs OpenWrt on an 650\,MHz MIPS CPU with 128\,MB RAM, weighs 73 grams, consumes less than 3 watts of power, measures 58$\times$58$\times$25\,mm, and is available from Amazon for less than \$US\,35 at the time of writing. The device features a dedicated hardware switch and programmable indicator lights which we have used to provide users with control over our access to their data (cf. Section \ref{sec:ethics}). When deployed as a sensor, collected data can be viewed immediately in the sensor's web interface, stored on an attached \ac{USB} device, or transmitted over the network.

\paragraph{Sensor Performance} 
When used in a man-in-the-middle attack setting, the router needs to intercept TR-069 traffic and simultaneously forward all other network communication. To evaluate its performance, we applied a constant load of 90\,Mbit/s forward traffic and then gradually added TR-069 clients simulating heavy TR-069 usage.
Figure~\ref{fig:sensorperf} shows the performance characteristics of the sensor relative to the number of connected TR-069 clients. The proxy can handle 45 plaintext or 20 \ac{TLS}-protected client sessions before significant increases in latency and ultimately request failures are observed. The use of \ac{TLS} induces an additional latency of 1100\,ms due to asymmetric cryptography operations on the device (we use 2048\,bit RSA keys in our tests). %
The router's price, size, power consumption and performance fulfill our (informal) requirements so that we can recommend it as a practical platform for TR-069 monitoring.

\section{Virtual Test Environment}
\label{sec:virtual-test-env}
\begin{figure}
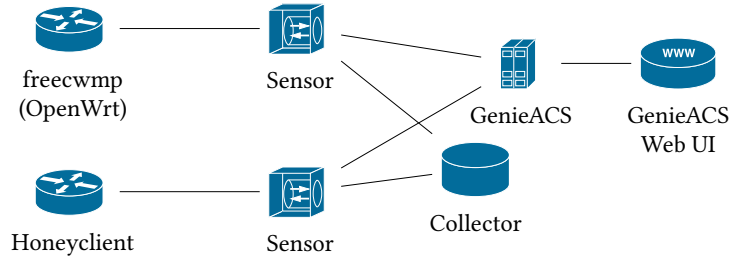

	\centering
	\includestandalone[page=7,width=.8\linewidth]{resources/illustrations}
	\caption{
		\label{fig:docker}
		Schema of the Docker-based test environment.
	}
\end{figure}
\noindent
One reason for the lack of TR-069 research could be that even basic inspection requires quite complex setups and the lack of an open platform for research represents a significant barrier \cite{sgx}.
While anyone can peek at their browser traffic by opening the browser's developer tools, TR-069 inspection typically requires a hardware device as client, a self-hosted configuration server to issue commands, and some form of monitoring software. To lower the entry barrier for TR-069 research and to test our monitoring software, we have built a network of Docker~\cite{docker} containers that emulates a full TR-069 environment on a single machine. 

Figure~\ref{fig:docker} depicts the basic network topology that can be spun up by invoking \texttt{docker-compose up}. The test environment consists of two TR-069 clients -- an OpenWrt machine running freecwmp and our honeyclient --, GenieACS as the configuration server, and our monitoring infrastructure in between. Interested users can control clients using GenieACS' web interface, manually interact with the configuration server using the honeyclient, view traffic on each sensor's web interface, or process the data recorded by the collector. We hope that our work is useful to other researchers starting to look at TR-069.

\section{Overview of TR-069 Remote Procedure Calls}
\label{sec:rpcs}
\newcommand{\cmdRow}[4]{#3 & #4 & #1 & #2 \\}
The following table lists TR-069 \acp{RPC} that are particularly interesting from a security and privacy perspective.\\[1ex]

\scriptsize
\noindent
\begin{tabularx}{\textwidth}{cclX}
	\toprule
	\hspace{-0cm}Client\hspace{-0cm} & \hspace{-0cm}Server\hspace{-0cm} & Name & Description\\
	\midrule
	\cmdRow{GetRPCMethods}{Obtain the list of commands accepted by the \ac{ACS} or client, including vendor-specific methods}{●}{●}	
	\cmdRow{Inform}{Inform \acs{ACS} about device and its state after connection establishment (see above).}{}{●}
	\cmdRow{Get-/SetParameterValues}{Read/write parameters on the client. For example, this can be used to set the wireless network SSID.}{●}{}
	\cmdRow{Get-/SetParameterAttributes}{ Read/write parameter attributes on the client. For example, this can be used to instruct the client to notify the \ac{ACS} when a parameter changes.}{●}{}
	\cmdRow{Add-/DeleteObject}{Modify parameter objects on the client. For example, this can be used to add a device with a static \ac{IP} address to the router configuration.}{●}{}
	\cmdRow{Download}{Download a file to the client and apply it. For example, this can be use to install new firmware.}{●}{}
	\cmdRow{Upload}{Upload a file from the client to the \ac{ACS}. For example, this can be used to upload diagnostic logs.}{❍}{}
	\cmdRow{ChangeDUState}{Install, update or uninstall additional software modules. For example, this can be used to enable additional functionality a customer has purchased.}{❍}{}
	\midrule
	\multicolumn{4}{p{.95\linewidth}}{
		\noindent
		● denotes commands that the endpoint must be ready to receive and process, \vfill ❍ denotes optional commands. \vfill See \cite[p. 74]{tr069} for a comprehensive list of commands.
	} \\\bottomrule
\end{tabularx}

\newpage
\section{Exemplary TR-069 Inform}
\label{lst:tr069example}
\footnotesize{
\begin{minted}{http}
POST / HTTP/1.1
Host: acs.example.com:7547
Authorization: Basic YmFzZTY0OiBpbnZhbGlkIGlucHV0Cg==
Content-Type: text/xml; charset="utf-8"

<cwmp:Inform>
<DeviceId>
	<Manufacturer>AVM</Manufacturer>
	<OUI>00040E</OUI>
	<SerialNumber>001DEAD3BEEF2</SerialNumber>
</DeviceId>
<Event>
	<EventCode>0 BOOTSTRAP</EventCode>
</Event>
<ParameterList>
	<Name>InternetGatewayDevice.DeviceSummary</Name>
	<Value>InternetGatewayDevice:1.4[](Baseline:2, EthernetLAN:1, ADSLWAN:1, Time:2, IPPing:1, WiFiLAN:2, DeviceAssociation:1), VoiceService:1.0[2](SIPEndpoint:1, Endpoint:1, TAEndpoint:1), StorageService:1.0[1](Baseline:1, FTPServer:1, NetServer:1, HTTPServer:1, UserAccess:1, VolumeConfig:1)</Value>
	<Name>InternetGatewayDevice.DeviceInfo.SoftwareVersion</Name>
	<Value>29.04.88</Value>
	<Name>InternetGatewayDevice.WANDevice.1.WANConnectionDevice.1.WANIPConnection.1.ExternalIPAddress</Name>
	<Value>10.0.0.4</Value>
</ParameterList>
</cwmp:Inform>
 
\end{minted}		
\begin{minted}{http}
HTTP/1.1 200 OK
Content-Type: text/xml; charset="utf-8"

<cwmp:InformResponse></cwmp:InformResponse>
\end{minted}
}
\noindent
Example of an \texttt{Inform} command after performing a factory reset. Messages are heavily condensed for brevity.
\newpage
\bibliographystyle{splncs04}
\bibliography{references}

\end{document}